# Evolution of Ge wetting layers growing on smooth and rough Si (001) surfaces: isolated {105} facets as a kinetic factor of stress relaxation

Larisa V. Arapkina, Kirill V. Chizh, Vladimir P. Dubkov, Mikhail S. Storozhevykh and Vladimir A. Yuryev

A. M. Prokhorov General Physics Institute of the Russian Academy of Sciences, 38 Vavilov Street, Moscow, 119991, Russia

Corresponding author:
Larisa V. Arapkina arapkina@kapella.gpi.ru

**Abstract**

The results of STM and RHEED studies of a thin Ge film grown on the Si/Si(001) epitaxial layers with different surface relief are presented. Process of the partial stress relaxation was accompanied by changes in the surface structure of the Ge wetting layer. Besides the well-known sequence of surface reconstructions ($2 \times 1 \rightarrow 2 \times N \rightarrow M \times N$ patches) and hut-clusters faceted with {105} planes, the formation of isolated {105} planes, which faceted the edges of $M \times N$ patches, has been observed owing to the deposition of Ge on a rough Si/Si (001) surface. A model of the isolated {105} facet formation has been proposed based on the assumption that the mutual arrangement of the monoatomic steps on the initial Si surface promotes the wetting layer formation with the inhomogeneously distributed thickness that results in the appearance of $M \times N$ patches partially surrounded by deeper trenches than those observed in the usual Ge wetting layer grown on the smooth Si(001) surface. Isolated {105} facets are an inherent part of the Ge wetting layer structure and their formation decreases the surface energy of the Ge wetting layer.

## 1. Introduction

The formation of quantum dots (QD) in the Ge/Si system has been studied for a long time [1–6]. To date, it is commonly adopted that self-organized Ge QDs are formed spontaneously in the Stranski-Krastanow growth mode due to the lattice mismatch of the Ge film and Si (001) substrate by 4.2 %. The shape and density of QDs depend on the growth temperature [4,5]. If a Ge QD array grows at low temperature (less than 550°C) it consists of hut clusters, i.e. those faceted by the {105} planes, namely, of pyramids (hut clusters with a square base) and wedges (hut clusters with a rectangular base) [1–9]. The growth of QD arrays occurs in two sequential stages: the first stage is the formation of a wetting layer (WL) and the second one is the appearance of QDs on the WL surface. The QDs nucleate only after the WL thickness exceeds some critical value depending on the growth temperature, the Ge contest in the $Si_{1-x}Ge_x$ WL and the deposition method [4,5,10]. On the basis of energetic consideration under conditions of thermodynamic equilibrium, it has been concluded that the driving force for the formation of QDs is relaxation of elastic stresses, which accumulate in the growing Ge film with the increase of its thickness. At the stage of wetting layer growth, the change in the total energy of the system might be considered as that in the WL surface energy and the strain energy accumulated in the WL [6]. The changes in the total energy occur due to the partial relaxation of elastic stresses and the decrease in the surface energy of the WL [11–20], which are manifested as the appearance of the surface reconstructions having lower surface energy under compressive stresses. In STM images, they have been observed as a sequence of the surface structures that appear with the increase of the WL thickness ($2 \times 1 \to 2 \times N \to M \times N$ patches separated by 1–2 monolayers (ML) deep trenches, the so-called dimer vacancy lines (DVL), and deeper ones, dimer row vacancies (DRV), often reaching the Si surface [21]);

this sequence of the WL alterations remains in a wide range of QD array growth temperatures [2–5,7,12,22–36]. After the WL thickness reaches its critical value, the continuation of the Ge deposition results in the increase in compressive stress in the growing Ge film and the strain relaxation (elastic stress release) occurs though the nucleation and growth of QDs on the WL surface [1,2,21,31,32,34,37–43] due to the gain in the reduction of the elastic energy compared to the increase in the surface energy [3–5,19,44,45], with the increase in the surface energy of the Ge film with the {105} planes arisen to facet QDs being due to increasing surface area.

In this paper, we present the results of the study of the changes in the Ge WL structure depending upon a relief of the initial Si/Si (001) surface. We have investigated the structure of WL grown on a rough surface of the Si/Si (001) epitaxial layer composed of large islands with a height of no more than 3 ML. Using STM, we have observed the formation of isolated {105} planes faceting the edges of $M \times N$ patches as well as QDs; the latter demonstrated a lower density and smaller sizes than they did when growing on a smooth Si/Si (001) surface, however.

We also discuss the issue whether the isolated {105} facets are one of the manifestations of a partial relaxation of elastic stresses in the Ge wetting layer, which occurs during the Ge film growth on a rough Si/Si (001) surface.

## 2. Experimental

The structures investigated in this work were obtained by depositing Si and Ge layers using molecular beam epitaxy (MBE) under ultra-high vacuum (UHV) conditions in an EVA 32 MBE chamber (Riber) connected with a GPI-300 (Sigma Scan) scanning tunneling microscope (STM) by UHV transfer line. The samples of $8 \times 8$ cm$^2$ in size were cut from commercial boron-doped CZ Si (001) wafers tilted ~0.2° towards the [001] direction ($\rho = 12$ Ω cm).

Before growth, the Si sample were subjected to a preliminary chemical treatment by the RCA etchant. After that, they were fixed in a molybdenum holders and loaded into a UHV cleaning chamber where they were outgassed at 600°C for 6 h. To decompose a SiO$_2$ layer and obtain a clean Si(001) surface, the samples were treated by a weak flux of Si atoms (the deposition rate was < 0.1 Å/s) at the temperature of 800°C in the MBE chamber [21,46]. Immediately after that, a 100-nm thick Si film was being grown at 650°C. The surface of this Si layer was used as the smooth initial one for the further growth processes.

For the preparation of the rough initial Si surface, the sample temperature was lowered to 360°C, at which a 50-nm thick Si film was deposited on the described above 100-nm smooth layer.

Thus, two types of the initial Si (001) samples with different surface reliefs were prepared for the further growth of a Ge film. Then, Ge layers with thickness less than 9Å were deposited at the temperature of either 360 or 650°C.

During all the processes, the samples were heated using tantalum radiation heaters installed at the rear side of a sample. The process temperature was monitored with a chromel–alumel and tungsten–rhenium thermocouples of the heaters in the cleaning and MBE chambers, respectively, which were *in situ* graduated beforehand against an IS12-Si infrared pyrometer (IMPAC). Before the processes, the MBE chamber was evacuated down to about $10^{-11}$ Torr. During the preliminary outgassing and growth, the pressure in the MBE chamber did not exceed $5\times10^{-9}$ Torr. The deposition rate and coverage were measured using XTC751-001-G1 film thickness monitors (Inficon Leybold-Heraeus) equipped with graduated in advance quartz sensors installed in the MBE chamber. Si and Ge were deposited from sources with the electron beam evaporation at the fluxes of ~0.3 and ~0.1 Å/s, respectively. The sample cooling down rate was ~0.4 °C/s after the preparation [8].

The composition of residual atmosphere in the MBE camber was monitored using the SRS RGA-200 residual gas analyzer (SRS) before and during the process.

The structure of the growth surface was monitored in situ during the processes with an RH20 reflected high energy electron diffraction (RHEED) tool (Staib Instruments). The STM measurements were carried out at room temperature using tungsten probe tips. The STM images were recorded in the constant current mode and processed using the WSxM software [47].

# 3. Results

## 3.1 Structure of the growth surface of the Si/Si (001).

Two types of the samples with thick Si films deposited at 650 and 360°C on the Si (001) substrates were prepared for the investigation of the influence of a surface relief on QD Ge growth. Fig.1(a) and (b) present STM images of the surfaces of these samples. It could be observed that the film growth at the high temperature (650°C) was in the step flow mode and led to the formation of a smooth surface consisting of terraces separated by monoatomic steps, the edges of which were arranged along the [001] direction [48]. The growth at the lower temperature (360°C) was in the island mode and resulted in the formation of a rough surface composed of islands. The corresponding RHEED images presented in Fig.2 support the STM results. The RHEED pattern for the smooth surface (Fig.2(a)) is seen to consist of narrow streaks corresponding to the $2 \times 1$ surface reconstruction, whereas the pattern for the rough surface (Fig.2(b)) is evidently formed by broad diffuse streaks with a faint 3D-related structure (discontinuous streaks) and also corresponds to the $2 \times 1$ reconstruction. Fig.1 (c) and (d) present the magnified STM images of the surfaces shown in (a) and (b) panels, respectively. The smooth surface of the Si film grown at 650°C (Fig.1(c)) is seen to be formed by dimer rows and has some defects: Si ad-atoms and vacancies marked as arrow 1 and arrow 2, respectively; arrow 3 points at the edge of the monoatomic step. The rough surface of the Si film grown at 360°C (Fig.1(d)) is mostly formed by stepped islands (macro islands), 3 ML high (1 ML ≈ 1.38 Å), with a wide flat top; some islands reach 4 ML height, however. The island base width is mostly over 20 nm and the average island side slope angles are from 0.5 to 2° that corresponds to the terrace width from 16 to 4 nm. The surfaces of the island top as well as those of the terraces composing island slopes consist of short dimer rows forming the $2 \times 1$ reconstruction.

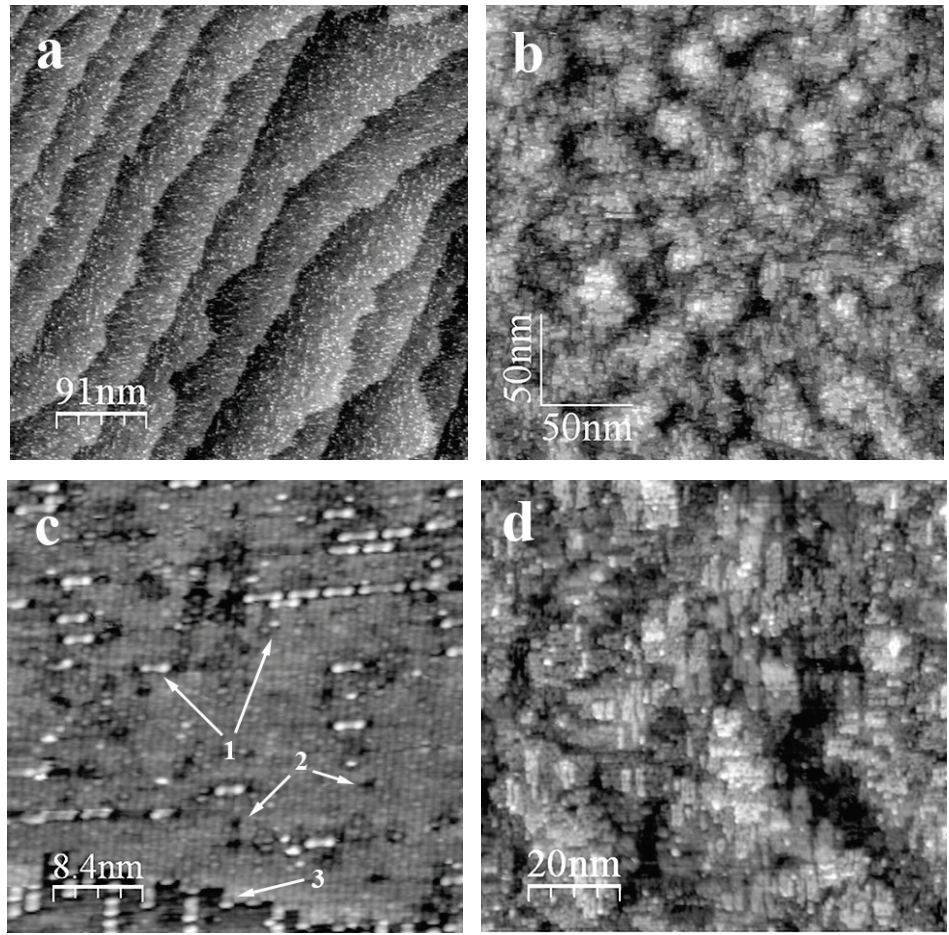

Fig.1. STM images of surfaces of the Si films grown at 650°C (a, c) and 360°C (b, d). (a) 455 × 455 nm, $U_t = +2.3$ V, $I_t = 0.1$ nA, the upper terraces are in the upper left corner of the frame; (b) 250 × 250 nm, $U_t = +2$ V, $I_t = 0.1$ nA; (c) 42 × 42 nm, $U_t = -2.8$ V, $I_t = 0.1$ nA, the upper terraces are in the top side of the frame; (d) 100 × 100 nm, $U_t = +2$ V, $I_t = 0.1$ nA; in the STM images, the <110> directions are almost parallel to the frame sides; the arrows point at Si ad-atoms (1), vacancies (2) and the edge of the monoatomic step (3).

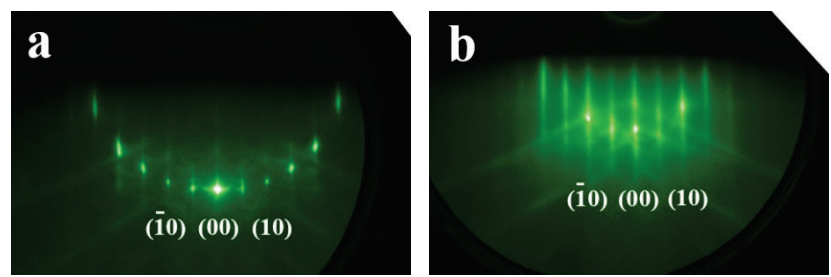

Fig.2. RHEED patterns of thick Si films deposited at 650 (a) and 360°C (b); [110] azimuth, the electron beam energy $E = 10$ keV.

## 3.2 Structure of thin Ge films

Ge layers were deposited at the coverage of 6 Å on the smooth and rough Si surfaces at 360°C. Fig.3 presents STM images of the film surfaces. As expected, Ge QDs formed as hut clusters faceted by {105} planes. The sizes and density of the Ge hut clusters were found to depend on the roughness of the initial Si layer. In Fig.3 (a) and (c), the Ge QD array grown on the smooth Si layer is shown. The cluster number density is seen to reach to ~$6\times10^{11}$ cm$^{-2}$, with the mean cluster height being from about 8 to about 9 Å (6 to 7 ML). The hut array formed on the rough Si surface (Fig.3(b) and (d)) shows the cluster number density of only ~$4\times10^{11}$ cm$^{-2}$; the hut mean height is also seen to be lower: it reaches about 7 Å (5 ML) in this case. The comparison of the Ge QD distribution on smooth and rough Si surfaces (Fig.3 (a) and (b)) leads one to the conclusion that it is controlled by the surface structure of the underlying Si layer. The important feature of the thin Ge film growth at 360°C is that the rough underlying Si layer is preserved and it is visible through the thin Ge film in STM images.

It was shown in our earlier work [32] that, during growth at 360°C, the thin wetting layers repeated the relief of the initial Si surface formed by monoatomic steps. One can observe in Fig.3 (a) that QDs are distributed randomly when grown on the smooth Si surface (Fig.1 (a) and (c)). In the case of the Ge film growth on the rough Si surface (Fig.1 (b) and (d)), large islands, which form the initial relief of the Si surface, remain. Thus, on a macro scale, the relief of the Ge wetting layer repeats that of the initial Si surface. Ge clusters prefer to arrange near slopes of large islands (Fig.3 (b)).

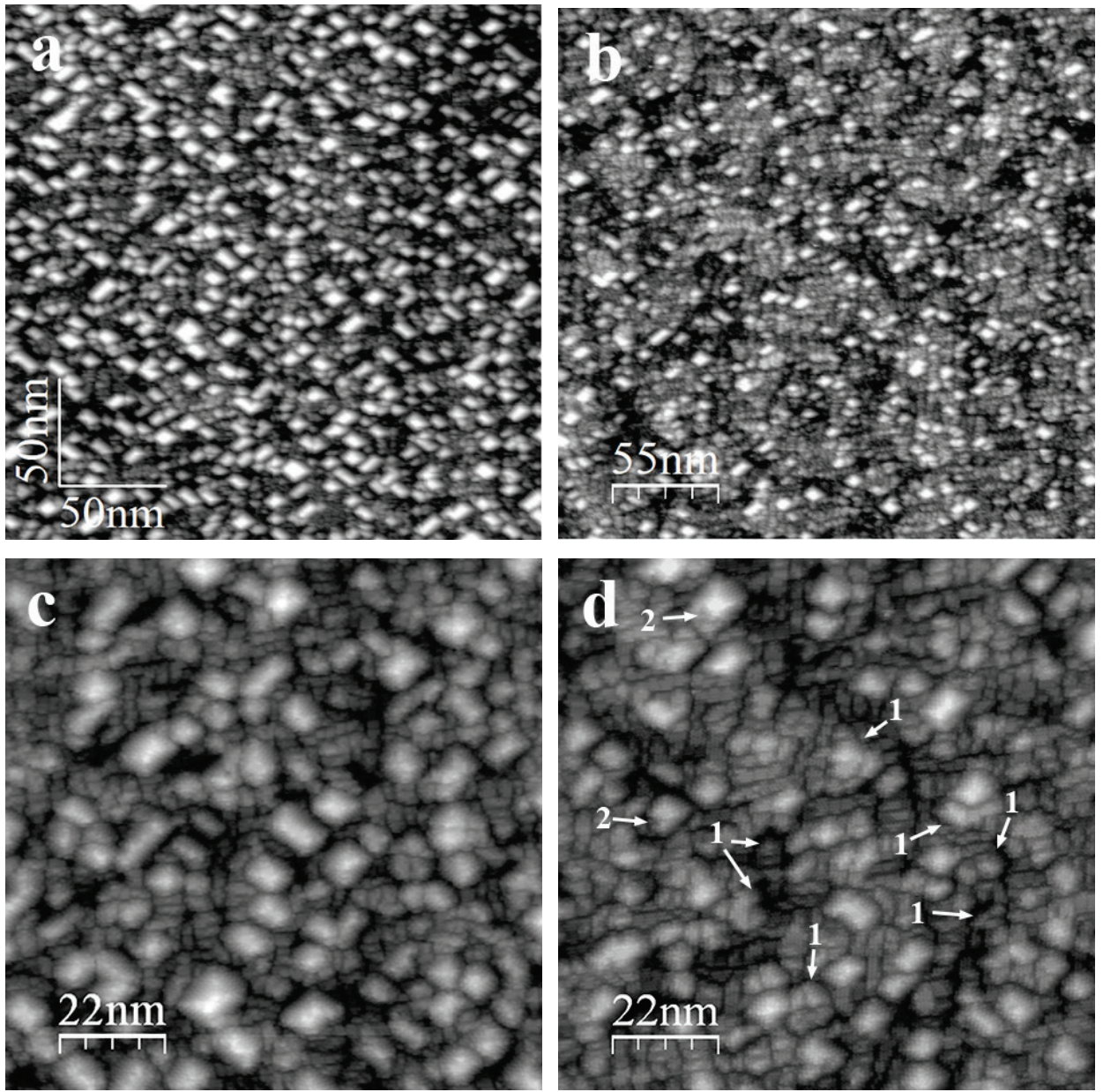

Fig.3. STM image of the surface of Ge films (the Ge coverage was 6 Å) deposited at 360°C on the Si layers grown at: (a) $T = 650°C$, 250 × 250 nm, $U_t = +2$ V, $I_t = 0.1$ nA; (b) $T = 360°C$, 272 × 272 nm, $U_t = +2$V, $I_t = 0.1$ nA; (c) $T = 650°C$, 110 × 110 nm, $U_t = -2$V, $I_t = 0.1$ nA; (d) $T = 360°C$, 110 × 110 nm, $U_t = -2$ V, $I_t = 0.1$ nA; the <110> directions are almost parallel to the frame edges in all the panels; the arrows point at the isolated {105} planes (1) and unfinished Ge clusters with 2 or 3 {105} facets (2).

It is also shown in Fig.3 (d) that, as distinct from the Ge QD array formed on the smooth Si surface, the Ge QD array formed on the rough Si surface demonstrates some peculiarities manifesting in the appearance of numerous isolated {105} planes (arrows 1 in Fig.3 (d)) and unfinished Ge clusters with 2 or 3 {105} facets (arrows 2). The corresponding magnified STM images presented in Fig.4. Fig.4 (a) shows the surface of a Ge film grown on the smooth Si surface. It is seen that hut clusters mostly have rectangular bases and are faceted by {105} planes, as it usually takes place if Ge QDs grow at low temperature [1,2,4–8,31,32,34,37–43]. Several isolated {105} planes are observed in these images as well (Fig.3 (c), Fig.4 (a)). In the case of the Ge QDs array grown on the rough initial Si surface (Fig.4 (b) and (c)), Ge huts have smaller sizes (the base width and therefore the height). (Incomplete Ge huts are marked with arrows 2 and the isolated {105} planes are pointed out by arrows 1 in Fig.4).

Panels (b) and (c) in Fig.4 demonstrate the STM images obtained at the empty-state mode (a negative voltage $U_t$ is applied to the STM tip) that allows one to count the number of MLs of WL (i.e. its thickness) involved in the formation of {105} planes. It is easy to be convinced that isolated {105} facets are formed by 3 or more MLs.

Comparing the Ge quantum dot arrays in Fig.3 (c) and (d) one can see the both surfaces to be not completely filled by Ge clusters and empty WL with a considerable unoccupied area is present on them. Empty WL covers a larger surface area when grown on the rough Si layer than on the smooth one. In Fig.3 (c) and Fig.4 (a) presenting images for WL on the smooth Si layer, WL is seen to be formed by $M \times N$ patches mainly separated by usual DVL and DRV trenches except for some points, likely lying over Si layer steps, at which deeper trenches form that reach the depth of 4 ML. Isolated {105} planes are seldom observed in this case.

In the case of the rough initial Si surface (Fig.3 (d) and Fig.4 (b), (c)), WL can be divided into areas of two kinds in accordance with the structure. First, there are large areas practically free of huts (see the middle of Fig.4 (c)), where WL consists of $M \times N$ patches separated by DVL and DRV trenches; the patches are similar to those presented in Fig.4 (a). The formation of isolated {105} planes was also rarely observed in these areas. These domains are associated with flat top areas of islets of the initial Si surface. Second, large areas of preferential growth of hut clusters (see

Fig.4 (b)), where $M \times N$ patches are often separated by deeper trenches up to 4 ML in depth. The formation of isolated {105} planes and unfinished hut clusters is more often observed in these domains. These areas arise over slopes of large Si islands composed of terraces and monatomic steps. In all the cases, the $c(4 \times 2)$ (arrows 3 in Fig.4) or $p(2 \times 2)$ (arrow 4 in Fig.4) reconstructions form on tops of patches, as it always takes place when WL grows at low temperatures [24,32,34,35,39,41].

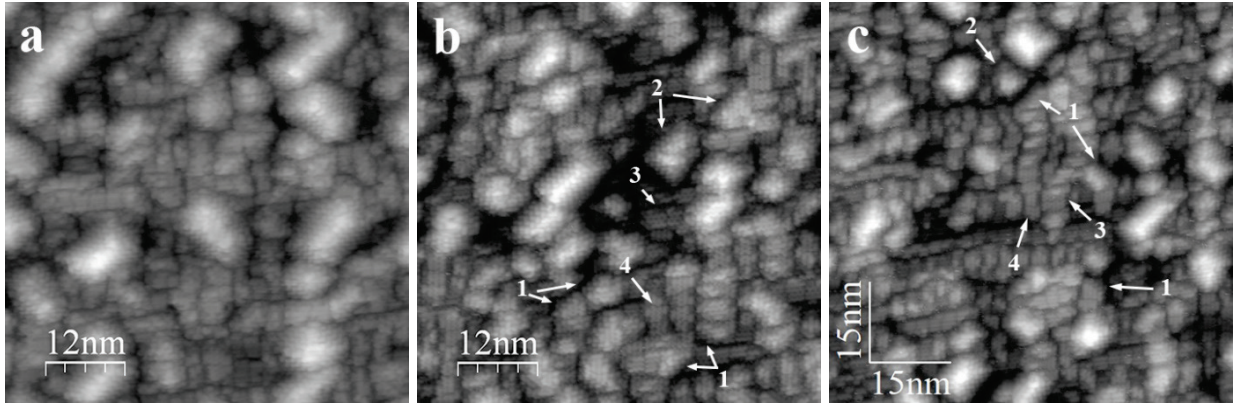

Fig.4. Magnified STM images of the surface of Ge films (6-Å Ge coverage) deposited at 360°C on the Si layers grown at 650 (a) and 360°C (b, c): (a) $60 \times 60$ nm, $U_t = -2$ V, $I_t = 0.1$ nA; (b) $62 \times 62$ nm, $U_t = -2$ V, $I_t = 0.1$ nA; (c) $77 \times 77$ nm, $U_t = -2$ V, $I_t = 0.1$ nA; in all the STM images, the <110> directions are almost parallel to the frame edges for all the panels; the arrows indicate the isolated {105} planes (1), unfinished Ge clusters with 2 or 3 {105} facets (2), and areas with the $c(4 \times 2)$ (3) and $p(2 \times 2)$ (4) reconstructions.

Fig.5 (a) shows a RHEED pattern of a Ge film deposited at the coverage of 6 Å and the temperature of 360°C on the smooth initial Si layer. As it is seen, this pattern corresponds to the beginning of the 3D-related structure formation — broken main streaks are apparent. Besides, the weak signals corresponding to the {105} planes begin appearing in the form of inclined streaks, which are marked by arrows in Fig.5 (a).

Fig.5 (b) demonstrates a RHEED pattern of the Ge film grown at the coverage of 6 Å and the temperature of 360°C on the rough initial Si layer. The diffraction pattern corresponds to that usually obtained for the WL surface consisting of $M \times N$ patches [28,49]. The 3D-related structure and streaks corresponding to the {105} planes are barely discernible although there are both Ge hut clusters and the isolated {105} planes on the surface (Fig.3 (b) and (d)). The change in the RHEED patterns therefore does not exactly coincide with the moment of the transition from the wetting layer to QDs [49].

Fig.5 (c) presents a RHEED pattern registered for the surface of the Ge film with the 9-Å coverage, the STM image of which is shown in Fig.6 (a). A pronounced 3D-related structure and strong streaks coming from {105} planes faceting the Ge quantum dots is observed in the pattern (the streaks are marked by arrows) [28,49].

Thus, the RHEED patterns presented in Fig.5 demonstrate the transition from the surface mostly coated by WL with tiny Ge QDs (Fig.5 (b)) to that covered by a dense QD array formed by larger QDs, which does not contain unoccupied areas of WL (Fig.5 (c)).

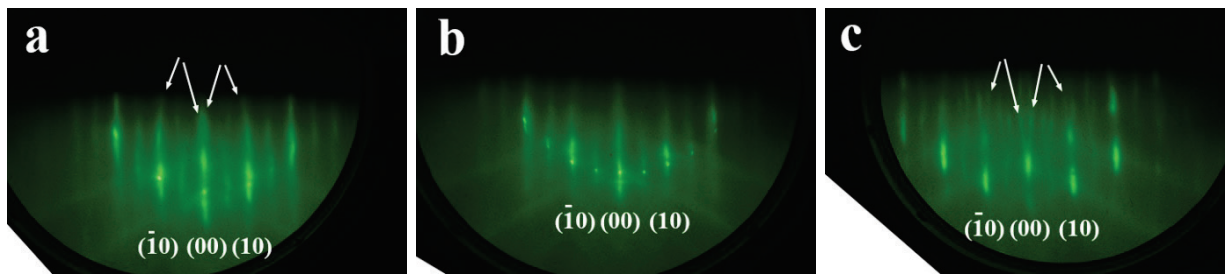

Fig.5. RHEED patterns of thin Ge films deposited at 360°C on the Si/Si(001) layers with different surface structure: Ge films with the coverage of 6 Å on the smooth (a) and rough (b) Si surfaces, and (c) a Ge film with the coverage of 9 Å on the smooth Si surface; [110] azimuth, the electron beam energy $E = 10$ keV; the arrows indicate the reflections from the {105} planes.

With the increase in the Ge coverage up to 9 Å, the hut clusters grown on the smooth Si layer increased their number density up to $\sim 8 \times 10^{11}$ cm$^{-2}$, whereas, on the rough initial Si layer, their number density reached $\sim 1 \times 10^{12}$ cm$^{-2}$. The STM images of the Ge QD arrays are presented in Fig.6. The significant increase in the QD density from $\sim 4 \times 10^{11}$ cm$^{-2}$ (Fig.3(b) and (d)) to $\sim 1 \times 10^{12}$ cm$^{-2}$ during the array growth on the rough Si layer occurs because this array contains QDs with smaller sizes in comparison with those grown on the smooth surface. Additionally, isolated {105} facets were present in abundance in WL at the early stage of the array growth on the rough surface, which than disappeared. This might also be an additional factor that accounts for the significant increase in the QD number density in the arrays grown on the rough Si layer.

In both growth processes Ge QDs contact each other that distorted the shapes of the bases. The appropriate RHEED image observed regardless of initial Si surface type is presented in Fig.5 (c).

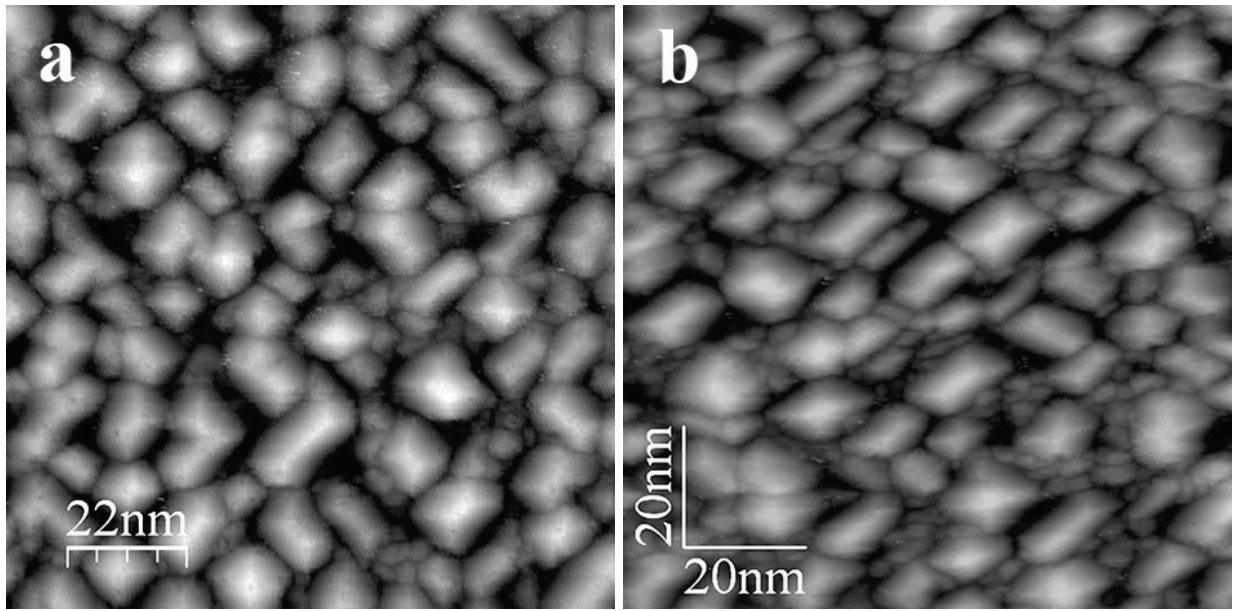

Fig.6. STM image of the surface of Ge films deposited at 360°C on the Si layers grown at smooth (a) and rough (b) Si surfaces, the Ge coverage was 9 Å: (a) 108 × 108 nm, $U_t = -2.3$ V, $I_t = 0.1$ nA; (b) 100 × 100 nm, $U_t = -1.8$ V, $I_t = 0.1$ nA; in all the images, the <110> directions are almost parallel to the frame edges.

In Fig.7, we present the results of the STM and RHEED study of a surface representing different structures formed during the WL growth. The Ge films with 4 Å and 6 Å coverage were grown on initial smooth Si layers. The STM image of the surface of the 4-Å Ge film grown at 650°C is demonstrated in Fig.7 (a). The high growth temperature enabled the formation of the smooth surface with clearly distinguishable monoatomic steps, which had been inherited from the Si(001) substrate. The upper step is in the top left corner of the STM image; the steps descend in the direction of the image bottom. The surface structure is 2 × N formed by DVLs running perpendicularly to the Ge dimer rows of the 2 × 1 reconstruction [4,50] that corresponds to the initial stage of the WL growth (1–2 ML) at a lower temperature. The RHEED pattern of this surface presented in Fig.7 (d) corresponds to the 2 × 1 reconstruction with additional narrow streaks emerging near the main ones (the former are indicated by the arrows); this corresponds to the 2 × N structure emerging on the Ge (001) surface due to the appearance of DVLs, i.e. shallow trenches formed due to the absence of every N-th dimer in rows in the [110] direction [4,50].

Fig.7 (b) demonstrates the STM image obtained at the surface of the 4-Å Ge film grown at 360°C. The monoatomic steps composing the initial Si (001) surface are observed after thin Ge film deposition. In the STM image, the upper step is in the left lower corner; the steps descend in the direction of the upper right corner of the frame. WL is composed of the 2 × N and M × N structures, which is caused by the formation of patches bounded by DRVs and DVLs [4,50]. As expected, Ge hut clusters are seen to be absent on this surface [32,34]. The RHEED pattern shown in Fig.7 (e) is similar to that presented in Fig.7 (d) but the weaker diffuse streaks marked by the arrows point out to the transition from the 2 × N reconstructed smooth surface to the rough one formed by M × N patches.

Fig.7 (c) shows the STM image of the Ge film with the 6-Å coverage grown in two stages. At the first stage, we deposited 4 Å of Ge at 650°C that led to the formation of the surface structure presented in Fig.7 (a); then at the second stage, we cooled down the sample to 360°C and deposited additional 2 Å of Ge. As a result, Ge hut clusters with the number density of ~$1.5 \times 10^{11}$ cm$^{-2}$ and the mean height of about 6 Å (~4 ML) arose on WL (compare with Refs. [9,32]). In addition, some amount of isolated {105} planes faceting deep pits formed in WL (marked by arrows 1) as well as unfinished hut clusters with {105} facets (arrows 2) appeared on the surface. The depth of faceted pits is seen to be 3 ML and more. WL is mostly formed by M × N patches surrounded by DRV and DVL trenches; they are similar to those observed in Fig.3 (c) and Fig.4 (a). In some surface areas, there are deeper trenches (over 2 ML in depth), near which isolated {105} facets and unfinished hut clusters are observed in most cases. $M \times N$ patches have been formed at the stage of the low-temperature growth and their height exceeds the thickness of the deposited 2 Å thick Ge film.

The RHEED pattern presented in Fig.7 (f) corresponds to the STM image in Fig.7 (c). As one can see, the signals of the $2 \times N$ structure have become very weak and the pattern has become similar to that of the 6-Å Ge film grown at 360°C on the rough Si surface (Fig.5 (b)).

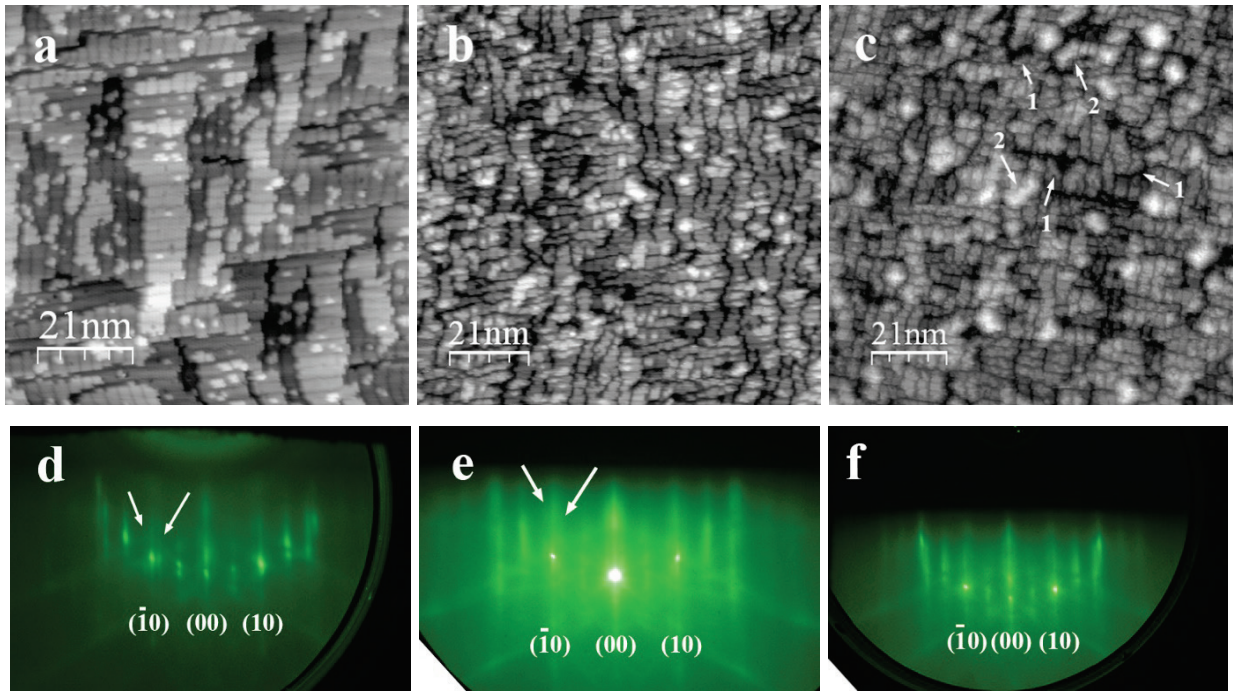

Fig.7. STM images and RHEED patterns of the surfaces of thin Ge films deposited on the smooth Si surfaces: (a, d) a 4-Å Ge film deposited at 650°C, 87 × 87 nm, $U_t = -1.8$ V, $I_t = 0.1$ nA; (b, e) a 4-Å Ge film deposited at 360°C, 102 × 102 nm, $U_t = -2$V, $I_t = 0.1$ nA; (c, f) a 4-Å Ge film deposited at 650°C coated by additional 2 Å of Ge deposited at 360°C, 110 × 110 nm, $U_t = -2.5$ V, $I_t = 0.12$ nA; the <110> directions are almost parallel to the frame edges in all the STM images; [110] azimuth, $E = 10$ keV in the RHEED patterns; the arrows in the panel (c) point at the isolated {105} planes (1) and unfinished hut clusters with {105} facets (2); the arrows in the panels (d) and (e) mark the streaks corresponding to the $2 \times N$ structure.

## 4. Discussion

The growth of thin films of Ge on silicon surfaces of different roughness has shown that it occurs in the Stranski-Krastanow mode and, besides expected hut clusters faceted by the {105} planes, isolated {105} planes form on sides of WL patches. With increasing Ge WL thickness, the compression stresses increase in WL that facilitates the stabilization of the {105} facets when WL is growing not only on flat but even on vicinal Si(001) substrates [3,14,15,36,51–56].

We suppose that a relief of the initial Si surface also has an influences on the WL structure and facilitates the appearance of isolated {105} planes. Our exploration of the thin Ge layers deposited at 360°C and 650°C has revealed that WL practically repeats the relief of the initial Si surface (Fig.1 (a) and (b)), formed by terraces separated by monoatomic steps, so the relief of the underlying layer is visible until the appearance of QDs (Fig.3 (a) and (b) and Fig.7 (b)). Ge atoms arriving on the growth surface from the molecular beam migrate on the terraces and attach to step edges. Just before hut cluster nucleation, WL surface is formed by $M \times N$ patches divided by trenches with the depth depending on the initial Si surface relief. For the smooth initial surface, they are ordinary DVLs and DRVs, whilst for the rough one, deeper trenches also form, which sometimes reach 4 ML in depth. We suppose that the formation of the deep trenches promotes the appearance of the isolated {105} facets.

At the WL formation on a rough initial Si surface composed of large islands (Fig.1 (b) and (c)), the same structural sequence from $2 \times 1$ to $M \times N$ patches took place as in the case of growth on the smooth Si surface. The difference is that WL has a structure formed by $M \times N$ patches surrounded by DVLs and DRVs in the areas of flat apexes of Si islands; however, in the regions nearby Si island slopes, Ge WL patches can also be bounded by deeper trenches exceeding 2 ML in depth.

Fig.8 presents the schemes of the WL evolution on smooth and rough initial Si surfaces under the assumption that the thickness of stable WL is only 3 ML (the left panels in Fig.8) and further increase in the Ge film thickness by 1 ML leads either to the elastic stress relaxation with the formation of QD nuclei [37] or to the appearance of isolated {105} planes (the right panels in Fig.8) in the WL structure. The model takes into account the influence of step edges on the local increase in the WL thickness.

The left panel of Fig.8 (a) shows a schematic of the sequential formation of a 3 ML thick Ge film for the case of the Ge deposition on the smooth initial Si surface consisting of wide terraces and monoatomic steps. The final surface of the Ge film (of a 3 ML thick Ge layer) consists of practically similar patches with the total thickness of 3 ML from the Ge/Si interface to the patch top, which are bounded by 2 ML deep trenches. The right panel of Fig.8 (a) shows a result of the deposition of the fourth ML of Ge that results in the elastic stress relaxation occurring through the nucleation of hut clusters faceted by four {105} planes; the hut nucleus is drawn as a white pyramid situated on a 3 ML thick patch of WL on the right panel of Fig.8 (a).

The left panel of Fig.8 (b) demonstrates the sequential formation of a 3 ML thick Ge film on the rough initial Si surface. The presence of a macro island on the initial Si surface causes the non-uniformity of the WL thickness within one patch. Layer segments of the excess thickness are highlighted in lighter gray and indicated with arrows in Fig.8 (b) (the left panel, 3 ML thick Ge layer). The total thickness of a patch at these segments is 4 ML. Thus, patches can end off with deeper trenches (3 ML deep) in the places above the step edges.

When depositing the next (the fourth) Ge layer, the total patch thickness begins to essentially exceed 3 ML on one of its sides, reaching 5 ML (the right panel of Fig.8 (b)). As a result, a thick enough patch side gives rise to the formation of an isolated {105} facet. The patch drawn on the right panel of Fig.8 (b) is thicker on one side than that shown on the right panel of Fig.8 (a) (5 ML vs 4 ML). This provides enough free space for side faceting by the {105} plane though the transformation of the patch cliff originally formed by [110] step edges. Note, that there is no increase in the surface area in this case. The formation of the isolated {105} planes having the lower surface energy than the {100} ones results in a gain in the surface energy of WL. This should move away the start of the quantum dot nucleation like it happens if WL grows at a higher temperature when the Si-Ge intermixing contributes to an increase in the thickness of WL reducing elastic stresses (Fig.7 (c)) [57–60].

According to the STM images, the isolated {105} facets form from at least 3 ML. Sometimes, they are observed in samples grown on the smooth Si surface, however (Fig.7 (c)). In the works [12,23,31,34,39], the STM images with isolated {105} facets formed in either deep trenches or large depressions of WL were demonstrated. Their heights always exceeded 3 ML. Faceted pits formed in the thick enough WL (~7 to 8 ML) by {105} planes were observed in Refs. [7,10]. In the works [22,61–63], the authors presented results of exploration of thin films of Ge grown on the vicinal Si substrates. It was shown that the existing tilt of the surface contributed to the morphological transition from the planar growth of Ge WL to the rough structures consisting of compressively strained {105} facets. This process is mainly driven by the lowering of the surface energy of the Ge layer. So, we suppose that of the isolated {105} facets are structural part of Ge wetting layer (similar to one of the types of reconstruction) and their formation is one of the ways to reduce the surface energy of the compressed Ge WL, which is observed only if there are the patches or pits with sufficient height or depth (> 3 ML).

The increase in the content of {105} facets (isolated ones and huts) in areas over slopes of the islands of the rough initial Si surface should be considered as a results of the appearance of $M \times N$ patches surrounded by trenches, which slow down the surface migration of Ge ad-atoms [64,65] and introduce the kinetic factors in the growth model. A stronger effect might be observed at the growth on a rougher surface. The deeper the trenches are the more is their influence upon the Ge ad-atom diffusivity, the decrease of which increases the probability of the nucleation of {105} planes due to a raise in the ad-atom content near the trenches [66–68].

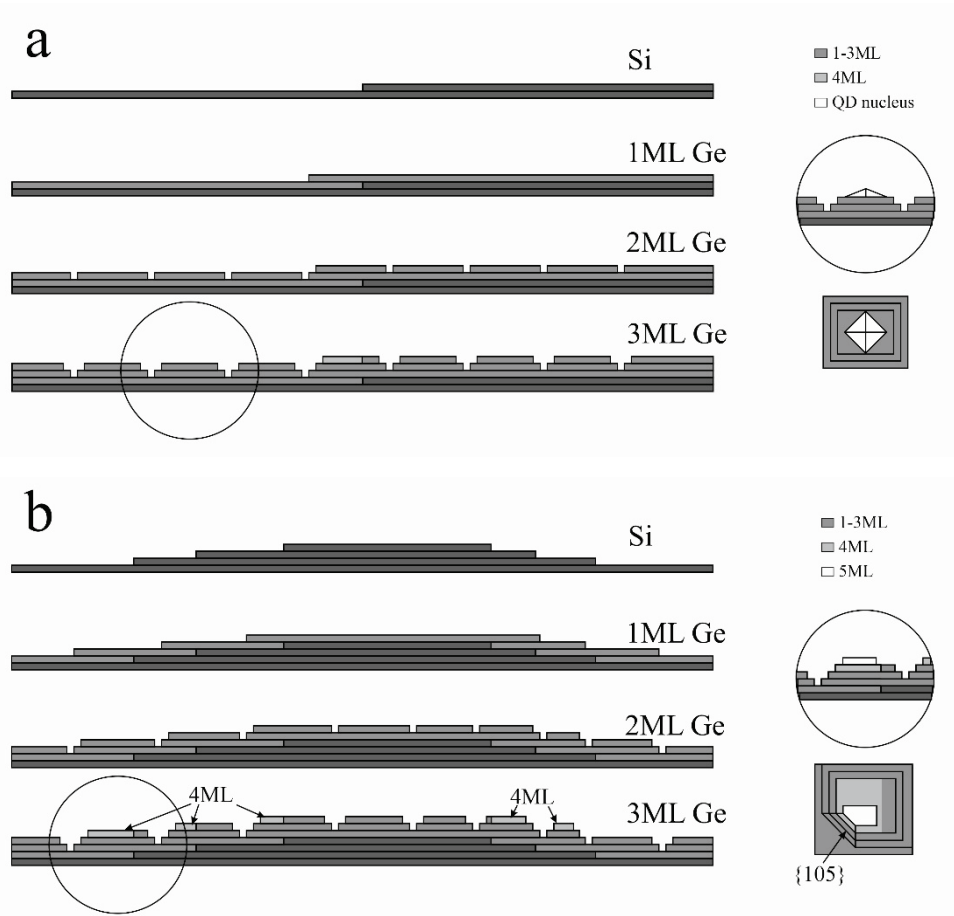

Fig.8. Schemes of the WL evolution on the smooth (a) and rough (b) initial Si surfaces: the left panels in (a) and (b) represent the layered deposition of a 3-ML Ge film; the right panels in (a) and (b) contain magnified images of the outlined WL domains with added 1 ML of Ge and their top views; the grayscale points the thickness of stable WL (1 to 3 ML) and added layers (4 ML, 5 ML or QD nucleus [37]).

## 5. Conclusion

The deposition of the thin films of Ge on the rough surface of the Si/Si (001) epitaxial layer composed of large islands with a height of no more than 3ML led to the formation of hut clusters faceted with {105} planes and isolated {105} facets. We interpret the latter as an inherent structural part of Ge wetting layer. The driving force of the process should be the lowering of the surface energy of Ge WL. We suppose the nucleation of isolated {105} facets can have an influence on the WL thickness critical value increasing it as long as the surface energy can be reduced.

The decrease in elastic energy becomes the main way to decrease the total energy after reaching the critical value of the WL thickness. With the increase in the WL thickness over the critical value, huts become the predominant type of Ge clusters, yet the isolated {105} facets are still observed until Ge QDs occupy the whole WL surface. We have proposed a scheme of formation of the isolated {105} facets based on the assumption that the mutual arrangement of monoatomic steps on the initial Si surface facilitates the formation of an inhomogeneous distribution of the WL thickness and the appearance of M×N patches partially surrounded by deeper trenches than they are in usual Ge WL grown on the smooth Si (001) surface. These deeper trenches are faceted by the {105} planes, which can reach the Si/Ge interface. The appearance of small-sized {105} facets in the form of either isolated ones or individual hut clusters cannot be definitely detected by the RHEED method.

**CRediT authorship contribution statement**

**Larisa Arapkina**: Investigation, Data Curation, Writing - Original Draft. **Kirill Chizh**: Investigation, Data Curation, Writing - Original Draft. **Mikhail Storozhevykh**: Investigation, Data Curation. **Vladimir Dubkov**: Investigation, Data Curation. **Vladimir Yuryev**: Data Curation, Writing - Review & Editing, Supervision.

**Declaration of Competing Interest**

The authors declare that they have no known competing financial interests or personal relationships that could have appeared to influence the work reported in this paper.

**Acknowledgments**

This research did not receive any specific grant from funding agencies in public, commercial or not-for-profit sectors. The Center for Collective Use of Scientific Equipment of GPI RAS supported this research via presenting admittance to its equipment.

27. B.Voigtlander, M.Kastner, Evolution of the strain relaxation in a Ge layer on Si(001) by reconstruction and intermixing, Phys. Rev. B 60 (1999) R5121-R5124. https://doi.org/10.1103/PhysRevB.60.R5121.

28. V.V.Dirko, K.A.Lozovoy, A.P.Kokhanenko, A.V.Voitsekhovskii, 2022. High-resolution RHEED analysis of dynamics of low-temperature superstructure transitions in Ge/Si(001) epitaxial system, Nanotechnology 33, 115603. https://doi.org/10.1088/1361-6528/ac3f56.

29. A.Vailionis, B.Cho, G.Glass, P.Desjardins, D.G.Cahill, J.E.Greene, Pathway for the Strain-Driven Two-Dimensional to Three-Dimensional Transition during Growth of Ge on Si(001). Phys. Rev. Lett. 85 (2000) 3672-3675. https://doi.org/10.1103/PhysRevLett.85.3672.

30. J.Tersoff, Missing dimers and strain relief in Ge films on Si(100), Phys. Rev. B 45 (1992) 8833-8836. https://doi.org/10.1103/PhysRevB.45.8833.

31. S.A.Teys, Different growth mechanisms of Ge by Stranski-Krastanow on Si (111) and (001) surfaces: An STM study, Applied Surface Science 392 (2017) 1017–1025. http://dx.doi.org/10.1016/j.apsusc.2016.09.124.

32. L.V.Arapkina, V.A.Yuryev, 2011. An initial phase of Ge hut array formation at low temperature on Si(001), J. Appl. Phys. 109, 104319. https://doi.org/10.1063/1.3592979.

33. L.V.Arapkina, V.A.Yuryev, 2012. Nucleation of Ge clusters at high temperatures on Ge/Si(001) wetting layer. J. Appl. Phys. 111, 094307. https://doi.org/10.1063/1.4707936.

34. L.V.Arapkina, V.A.Yuryev, 2013. On atomic structure of Ge huts growing on the Ge/Si(001) wetting layer, J. Appl. Phys. 114, 104304. https://doi.org/10.1063/1.4819457.

35. M.S.Storozhevykh, L.V.Arapkina, V.A.Yuryev, 2015. Evidence for Kinetic Limitations as a Controlling Factor of Ge Pyramid Formation: a Study of Structural Features of Ge/Si(001) Wetting Layer Formed by Ge Deposition at Room Temperature Followed by Annealing at 600°C, Nanoscale Research Letters 10, 295. http://doi.org/10.1186/s11671-015-0994-0.
23